\documentclass{article}

\def\IR{\mathchoice{ \hbox{${\rm I}\!{\rm R}$} }
                   { \hbox{${\rm I}\!{\rm R}$} }
                   { \hbox{$ \scriptstyle  {\rm I}\!{\rm R}$} }
                   { \hbox{$ \scriptscriptstyle  {\rm I}\!{\rm R}$}}}

\def\IP{\mathchoice{ \hbox{${\rm I}\!{\rm P}$} }
                   { \hbox{${\rm I}\!{\rm P}$} }
                   { \hbox{$ \scriptstyle  {\rm I}\!{\rm P}$} }
                   { \hbox{$ \scriptscriptstyle  {\rm I}\!{\rm P}$}}}

\def\IE{\mathchoice{ \hbox{${\rm I}\!{\rm E}$} }
                   { \hbox{${\rm I}\!{\rm E}$} }
                   { \hbox{$ \scriptstyle  {\rm I}\!{\rm E}$} }
                   { \hbox{$ \scriptscriptstyle  {\rm I}\!{\rm E}$}}}

\begin{document}

\title{Scattering theory from microscopic first principles}

\author{Detlef D\"urr \\ Mathematisches Institut der Universit\"at M\"unchen,\\ 
Theresienstr.\ 39,  
80333 M\"unchen, Germany \\ \\
Sheldon Goldstein \\ Department of Mathematics, Hill Center,
Rutgers: \\ The State University of New Jersey,\\
110 Frelinghuysen Road,
Piscataway, NJ 08854-8019\\ email: {\tt oldstein@math.rutgers.edu} \\ \\
Stefan Teufel \\ Zentrum Mathematik, Technische Universit\"at M\"unchen,\\ Gabelsbergerstr.\ 49,
80290 M\"unchen, Germany \\ \\
Nino Zangh\`{\i} \\ Dipartimento di Fisica, Universit\`a di Genova, Sezione INFN\\
Genova,   Via Dodecaneso 33, 16146 Genova, Italy}

\maketitle


\begin{abstract}
We sketch a derivation of abstract scattering theory from the microscopic
first principles defined by Bohmian mechanics. We emphasize the importance
of the flux-across-surfaces theorem for the derivation, and of randomness
in the impact parameter of the initial wave function---even for an,
inevitably inadequate, orthodox derivation.   
\end{abstract}

\maketitle

\thanks{Dedicated to Joel Lebowitz, with love and admiration,
for his 70th birthday. Supported in part by the DFG, by NSF Grant
No. DMS95--04556, and by the INFN.}

\maketitle

\section{Introduction}\label{intro}

Abstract scattering theory, or the $S$-matrix formalism, can be regarded as
a phenomenological description analogous to thermodynamics. And like
thermodynamics, it should be derivable from microscopic first
principles. It is somewhat surprising that while this was done long ago for
thermodynamics, by Boltzmann and Gibbs using the methods of statistical
mechanics, it has not yet been achieved for quantum scattering theory.

We believe there are two main sources of difficulty: (1) failure to pay
sufficiently careful attention to the experimental conditions in scattering
phenomena, and in particular to the fact that randomness in the initial
wave function is an experimental reality that is crucial to an
understanding of the emergence of the textbook formula for the differential
cross section, involving the absolute square of the momentum matrix elements
of the $T$-matrix; and (2) failure to pay sufficiently careful attention to
precisely which microscopic first principles the derivation could
conceivably be based upon. 

We shall argue that while orthodox quantum theory is not up to the job,
Bohmian mechanics is, and we shall sketch the derivation. Since scattering
theory is at the heart of the experimental evidence for quantum theory, we
believe that understanding how the formulas of scattering theory emerge
from microscopic first principles should be of general interest.

\section{The $S$-matrix}\label{S}

 The basic formula of abstract scattering theory concerns the
probability of finding a system in the {\em free state} $g$ asymptotically
in the future given that it was in the {\em free state} $f$ asymptotically
in the past. This is expressed in terms of the basic object of scattering theory,
the scattering operator $S$, usually called the $S$-matrix. The probability
$P(f\to g)$ for scattering from state $f$ to state $g$ is given by
\begin{equation} \label{S1}
P(f\to g) = |\langle g, S f\rangle|^2 \,,
\end{equation}
where $f$ and $g$ are members of some Hilbert space $\mathcal H$, the space
of {\em free} states, with inner product $\langle \cdot,\cdot \rangle$.

This formula is often considered very appealing since it makes no reference
to space-time processes, but directly relates experimental procedures:
``preparation'' in the distant past to ``measurement'' in the distant
future. From (\ref{S1}) one computes, via formal manipulations, values for
the experimentally relevant cross section, an issue which we shall take up
in Section~\ref{scs}.

We first review how expression (\ref{S1}) is understood in mathematical
physics as emerging from Hamiltonian quantum mechanics.

\section{The Schr\"odinger evolution and the $S$-matrix}\label{sc}

We shall be concerned here with the scattering of a single spinless quantum
particle off of a ``target,'' or, what amounts mathematically to more or
less the same thing, of a pair of spinless particles off of each
other.\footnote{Recall that the scattering of two particles interacting via
a translation invariant pair potential can be reduced to potential
scattering of one particle by a change of variables to relative and
center-of-mass coordinates. However, in quantum mechanics this is not as
trivial as in classical mechanics, since one also must assume for this that
the wave function is a product wave function in the new coordinates. This
will not be the case in general, but one can easily convince oneself that
this condition is satisfied, for example, in the case of two particles both
described by plane waves.}  We thus begin our analysis with the
non-relativistic quantum mechanics for a single spinless particle in an
external potential $V$.  

The state of the system at time $t$ is given by its wave function
$\psi_t\in L^2(\IR^3)$, which evolves according to Schr\"odinger's
equation
\begin{equation}\label{Schroedinger}
{\rm i}\frac{\partial \psi_t}{\partial t} = H\psi_t\,,
\end{equation} 
where $H = H_0 +V$ with $H_0 = -\frac{1}{2}\Delta$ 
(in units for which $\hbar=1$ and $m=1$). A solution $\psi_t$ is determined by a choice of 
the initial condition $\psi=\psi_0$ at time $t=0$, 
\begin{equation}\label{scatpsi}
\psi_t = {\rm e}^{-{\rm i}Ht}\psi_0\,.
\end{equation}
If the scattering potential $V$ decays sufficiently rapidly at spatial
infinity, one expects scattering states, i.e., states that eventually leave
the influence of the potential, to evolve for large positive times
according to the free dynamics given by $H_0$, i.e., that the motion is
asymptotically free. In the following definition this free motion,
defining the asymptotics, is invoked. We demand that for every scattering
state $\psi$ there exists a state $\psi_{{\rm out}}$ such that
\begin{equation}\label{free}
\lim_{t\to\infty}\|{\rm e}^{-{\rm i}Ht}\psi - {\rm e}^{-{\rm i}H_0t}\psi_{{\rm out}}\|=0\,.
\end{equation}
Thus, one is interested in the existence and the range of the 
wave operator 
\begin{equation}\label{O+}
\Omega_+ := \lim_{t\to\infty}{\rm e}^{{\rm i}Ht}{\rm e}^{-{\rm i}H_0t}\,,
\end{equation}
where the limit is in the strong sense.  If the wave operator
exists,\footnote{\label{foot}Note that it might appear physically natural
to define the wave operator as the inverse of $\Omega_+$, i.e., as the map
from scattering states $\psi$ to the corresponding future asymptotic states
$\psi_{{\rm out}}$. However, one does not know a priori which states are
scattering states. Thus the domain of definition of that operator would be
far from clear! In fact, the goal of the mathematical physics of scattering
theory is precisely to clarify such issues. With the definition (\ref{O+}),
this question is shifted to that of the range of $\Omega_+$.} every state
in its range eventually moves freely in the sense of (\ref{free}), since
$\Omega_+$ maps every ``free state'' $\psi_{{\rm out}}$ to the
corresponding ``scattering state'' $\psi$. One can repeat these
considerations for the behavior of wave functions in the distant past and
define analogously the wave operator
\begin{equation}
\Omega_- := \lim_{t\to-\infty}{\rm e}^{{\rm i}Ht}{\rm e}^{-{\rm i}H_0t}\,.
\end{equation}
It is well known and not difficult to see that the wave operators exist for
short-range potentials.\footnote{Short-range potentials basically decay, as
$x\to\infty$, like $|x|^{-1-\epsilon}$ for some $\epsilon>0$. In the case
of long-range potentials one must use, instead of ${\rm e}^{-{\rm i}H_0t}$,
a modified free dynamics to define the wave operators.}

Whenever the wave operators exist, they obey the intertwining relations,
which follow from a simple calculation:
\begin{equation}\label{inter1}
 {\rm e}^{-{\rm i}Ht}\Omega_\pm = \Omega_\pm {\rm e}^{-{\rm i}H_0t}\,.
\end{equation}
And thus, on the domain $D(H_0)$ of $H_0$, we have by differentiation
\begin{equation}\label{inter2}
H\Omega_\pm = \Omega_\pm H_0\,.
\end{equation}
As a consequence of this relation and the fact that $\Omega_\pm$ are
partial isometries (i.e., that they act unitarily from their domain to their
ranges ${\rm Ran}(\Omega_\pm)$ ) one concludes that the restrictions of $H$
to ${\rm Ran}(\Omega_\pm)$ are unitarily equivalent to $H_0$. As such, they
have the same spectrum, and we may conclude that ${\rm
Ran}(\Omega_\pm)\subset {\mathcal H}_{\rm ac}(H)$, the absolutely
continuous subspace of $H$, the set of all states having an absolutely
continuous spectral measure for$H$. Thus scattering states are very much
related to spectral theory.

As we remarked in footnote \ref{foot}, the task of determining the range of
the wave operators is less simple. It was one of the main preoccupations of
mathema\-tical scattering theory for several decades. From a physical point
of view one might expect that every state orthogonal to all bound states
eventually leaves the influence of the potential and moves freely, and hence
is in the range of the wave operators. Since the set of bound
states of $H$ is ${\mathcal H}_{\rm pp}(H)$, the spectral subspace of $H$
spanned by its eigenvectors, this is mathematically expressed by
\begin{equation}\label{AC}
{\rm Ran}(\Omega_{\pm}) = {\mathcal H}_{\rm cont}(H)\,,
\end{equation}
where ${\mathcal H}= {\mathcal H}_{\rm pp}(H) \oplus{\mathcal H}_{\rm
cont}(H)$. Wave operators (and the corresponding Hamiltonians $H$)
satisfying (\ref{AC}) are called {\it asymptotically complete}. When $H$ is
asymptotically complete the set of scattering states is precisely
${\mathcal H}_{\rm cont}$. Asymptotic completeness has been established for
many different systems, including many-particle systems (see, e.g.,
\cite{Enss,Dere,SigalSoffer} and the references therein).

The continuous part of the spectrum can in general be separated into two
parts, the absolutely continuous part, supporting spectral measures
absolutely continuous with respect to Lebesgue measure, and the singular
continuous part, supporting singular continuous spectral measures. With
what we already know from the existence of the wave operators, we may
conclude that a  Hamiltonian which is asymptotically complete has
no singular continuous spectrum.

Assuming asymptotic completeness, as we shall for the rest of this paper,
we turn to the standard description of the scattering experiment. A
scattering state is a solution of (\ref{scatpsi}) with $\psi\in{\mathcal
H}_{\rm ac}$ and with $t=0$ any time between preparation and detection. The
preparation is done at a very large negative time and the detection at a
very large positive time. The scattering state is expressed in terms of its
asymptotic in-state $\psi_{\rm in}: = \Omega_-^{-1}\psi\ (=f)$, which is
mapped by the scattering operator $S$ to the asymptotic out-state
$\psi_{\rm out}: = \Omega_+^{-1} \psi = S\psi_{\rm in}$, so that
\begin{equation}\label{Sdef} S:= \Omega_+^{-1}  \Omega_-\,.
\end{equation}
Since $\Omega_-: L^2(\IR^3) \to {\mathcal H}_{\rm ac}(H)$ and
$\Omega_+^{-1}:{\mathcal H}_{\rm ac}(H) \to L^2(\IR^3)$, the scattering
operator $S$ is well defined. In view of (\ref{free}), the scattering state at
the time of detection is close to $\psi_{\rm out}$ evolved forward in time
via the free evolution and at the time of preparation it is close to
$\psi_{\rm in}$ evolved backwards in time.

\section{The scattering cross section and the scattering process}\label{scs}

Textbook scattering theory is primarily concerned with transitions between
plane waves, states of well defined momentum, and this also seems to be of
primary interest to experimentalists. Roughly speaking, one tries to apply
equation (\ref{S1}) with $f$ and $g$ momentum eigenstates. For a variety of
reasons, this leads to many difficulties, some associated with the outgoing
state (or the out-process) and some with the incoming state (the
in-process).  The treatment of outgoing plane waves is superficially
straightforward from an orthodox perspective, and we shall focus in this
section primarily on coping with the in-process. Later, in
Sections~\ref{sic}--\ref{bcs}, we shall argue that even with regard to the
out-process, things are not as straightforward as they seem, that the
framework of orthodox quantum theory does not, in fact, provide an adequate
microscopic basis for scattering theory, and that Bohmian mechanics does.

Probabilities for transitions to plane waves correspond to the statistics
for the results of a final momentum measurement. In abstract scattering
theory, the scattering cross section is calculated as the probability that
the momentum of the asymptotic state in the far future lies in the cone
$C_\Sigma := \{k\in\IR^3: k/|k|\in\Sigma\}$, $\Sigma\subset S^2$, the
unit sphere in $\IR^3$. We shall assume that $\Sigma$ is
closed. According to the standard measurement formalism one integrates the
modulus square of the Fourier transform of the state at the time of
measurement (the momentum distribution) over the cone $C_\Sigma$. Since the
state at a large time $\tau$ is approximately ${\rm e}^{-{\rm i}H_0 \tau}
S\psi_{\rm in}$ and the momentum is preserved by the free evolution, the
relevant probability density is $|\langle k|S\psi_{\rm in}\rangle|^2=
|\widehat{S\psi_{\rm in}}(k)|^2$.  Thus the scattering cross section is given,
independently of $\tau$, by
\begin{equation}\label{cross}
\sigma^{\psi}(\Sigma) 
:=  \int_{C_\Sigma} |\widehat{\Omega_+^{-1}\psi}(k)|^2 \,{\rm d}^3k =
\int_{C_\Sigma} |\widehat{S\psi_{\rm in}}(k)|^2 \,{\rm d}^3k 
\end{equation}
for any scattering state $\psi$. This is the central formula of scattering
theory.

Since in scattering theory one is interested  in the {\it changes} that occur
during the scattering process, it is convenient to replace $S$ in
(\ref{cross}) by $T := S- I$.  We thus define
\begin{equation}\label{T} 
\sigma_T^\psi(\Sigma) :=  \int_{C_\Sigma} \left|\widehat{T\psi}_{\rm in}(k)\right|^2 \, {\rm d}^3k \,.
\end{equation} 
For the case in which $\psi_{\rm in}$ is an (approximate) plane wave,
$\sigma_T$ corresponds to the genuine scattering events, in which a change
in direction is detected; because most of the plane wave will never overlap
the scattering region, these occur only rarely in this case.  A
(heuristically) straightforward computation yields that $T$ is an integral
operator with kernel $-2\pi {\rm i}\delta(k^2/2 - k^{'2}/2)T(k,k')$, so
that
\begin{equation}\label{TT}
\widehat{T\psi}_{\rm in}(k)= -2\pi {\rm i}\int\limits_{|k'|=|k|} T(k,k')
\widehat\psi_{\rm in} (k') |k'| {\rm d}\Omega(k')
\end{equation}

We turn now to the in-process, the treatment of incoming plane waves. If we
substitute a plane wave for $\psi_{\rm in}$ in (\ref{T}), we obtain an
infinite quantity, proportional to $\delta(0)$. This is not terribly
astonishing since a plane wave is nonnormalizable and nonphysical.  A plane
wave is not a possible quantum state for a single particle. Rather, a plane
wave is often regarded as describing a spatially homogeneous beam of particles.

Moreover, it is with a prepared beam of particles, of approximate momentum
$k_0$, approximately spatially homogeneous prior to its reaching the
scattering region, that real-world scattering experiments are mainly
concerned. And the quantity of primary physical interest is such
experiments is the differential cross section  $\sigma_{\rm 
diff}^{k_0}(\Sigma)$, describing the rate at which   particles are
scattered into (i.e., measured in)  the solid angle $\Sigma$ when the  beam
has unit current (one particle per unit of time per unit of cross section area perpendicular to the beam). 

The infinite quantity obtained from (\ref{T}) by setting $\psi_{\rm in}
\sim {\rm e}^{{\rm i}k_0\cdot x}$ must be suitably normalized to obtain the
differential cross section.  A theoretical physics type argument in which
this is done can be found in \cite{BjorkenDrell,Low}. Very loosely
speaking, it is argued that by dividing with the quantum flux of the plane
wave through a unit area integrated over all time, another infinite
quantity, one cancels the $\delta(0)$ factor. It is claimed that the
computation yields
\begin{equation}\label{diff}
\sigma_{\rm diff}^{k_0}(\Sigma)= 16\pi^4\int_\Sigma |T(\omega|k_0|, k_0) |^2\,{\rm d}\Omega\,.
\end{equation} 
This formula---which is also suggested by naive scattering theory, see
Section~\ref{nst}---is, as we shall argue, correct. But the argument in
\cite{BjorkenDrell,Low} is, too say the least, somewhat
obscure. Moreover, even if it were in a sense crystal clear, it could not,
as we shall also explain, be regarded as providing a derivation of
(\ref{diff}) from microscopic first principles.

The point is that to the extent that the individual quantum particles in a
beam have a wave function at all, that wave function must be normalizable,
i.e., an element of the Hilbert space, and cannot be a plane
wave.\footnote{If the particles were in an entangled state, for example
because of symmetry, then the individual particles would not described by a
wave function at all. We shall assume here that we are dealing with
situations for which this possibility can be ignored.} Rather, the
particles in our homogeneous beam should be regarded as being, initially,
at time $-\tau$, in approximate momentum eigenstates, described by wave
functions $\psi_{-\tau}$ whose Fourier transform is supported in a small
neighborhood of $k_0$, $|\widehat\psi_{-\tau}(k)|^2 \approx \delta(k-k_0)$.
We must thus consider the limit in which the prepared wave functions, while
remaining normalized, achieves zero momentum spread:
$|\widehat\psi_{-\tau}(k)|^2 \to \delta(k-k_0)$.

The simplest way to model such a homogeneous beam is as follows: We
consider as input a spatially homogeneous collection of particles,
statistically and quantum mechanically independent and noninteracting (with
each other), moving with momentum $\approx k_0$ , where all particles have
at preparation wave functions identical up to translation: the prepared
wave functions are translates of a common wave function $\phi$ with
$|\widehat\phi(k)|^2 \approx \delta(k-k_0)$.  In such a beam the
``centers'' of the prepared wave functions are independently and uniformly
distributed in a plane perpendicular to $k_0$, far from the scattering
region and on the incoming side. More precisely, we model the beam by a
Poisson system of points $(y,t)$ corresponding to wave functions which are
prepared at a rate uniform in time and with centers $y$ uniformly
distributed in a two dimensional plane $\Gamma_L=\{-L \frac{k_0}{|k_0|} +
a\,\vert\, a\perp k_0\}$. The point $(y,t)$ corresponds to a particle whose
wave function at time $t\ (=-\tau)$ is $\phi_y$, where the subscript
indicates translation: $\phi_y$ is the translation of $\phi$ by $y$.  If,
as we shall assume, the Poisson system has unit density or intensity, then
the beam it describes has unit current.

Since each particle $(y,t)$ in the beam scatters into $\Sigma$ with
probability given by (\ref{T}) with $\psi_{\rm in}$ replaced by $\psi_{\rm
in}^y$, the in-state corresponding to $\phi_y$,\footnote{More precisely,
$\psi_{\rm in}^y= \Omega_-^{-1}\phi_y$, the in-state corresponding to
$(y,0)$. Clearly, by time-translation invariance, the scattering
probability is independent of $t$. This corresponds to the fact that the
in-state associated with $(y,t)$ is $e^{-iH_0t}\psi_{\rm in}^y$; the
outgoing momentum distribution corresponding to $(y,t)$ is thus independent
of $t$, since the free evolution commutes with $S$.} it follows that the
rate at which the particles of the beam scatter into $\Sigma$ is given by
the integral of this over the plane $\Gamma_L$. Since in the limit
$|\widehat\phi(k)|^2 \to \delta(k-k_0)$ the $\phi_y$'s will spread over the
scattering region, we must first perform the limit $L\to\infty$. We thus
obtain as the quantity that should yield the theoretical differential cross
section

\begin{equation}\label{dT}
\sigma_{\rm diff}^{k_0}(\Sigma)=\lim_{|\widehat\phi(k)|^2 \to \delta(k-k_0)}\lim_{L\to \infty}\int_{C_{\Sigma}}\int_{y\in\Gamma_L}
\left|\widehat{T\psi^y_{\rm in}}(k)\right|^2 \,{\rm d}^2y\, {\rm d}^3k\,,
\end{equation}
or, somewhat more explicitly, 
\begin{equation} \label{dO}  
\sigma_{\rm diff}^{k_0}(\Sigma)=\lim_{|\widehat\phi(k)|^2 \Rightarrow \delta(k-k_0)}\lim_{L\to \infty}\int_{C_{\Sigma}}\int_{y\in\Gamma_L}
\left|\widehat{\Omega_+^{-1}\phi_y}(k)\right|^2 \,{\rm d}^2y\, {\rm d}^3k\,,
\end{equation}
provided $k_0\notin C_{\Sigma}$.\footnote{If $V$ has bound states, $\phi_y$
typically will not be in ${\mathcal H}_{\rm ac}$. In this case, $\phi_y$ in
(\ref{dO}) should be replaced by $P_{{\mathcal H}_{\rm ac}}\phi_y$ and
$\psi^y_{\rm in}$ in (\ref{dT}) by $\Omega_-^{-1}P_{{\mathcal H}_{\rm
ac}}\phi_y$. The analysis sketched here would then have to be replaced by a
somewhat more complicated one. We ignore this possibility here.} The $\Rightarrow$ in (\ref{dO}) means that
the limit is such that $\widehat\phi(k)$ is {\it strictly} supported on a
neighborhood of $k_0$ that shrinks to $k_0$ (which is perhaps unrealistic
as an assumption on the prepared state). (\ref{dT}) and (\ref{dO}) need not
agree, even for $k_0 \notin C_\Sigma$, if the first limit in (\ref{dO})
were understood as allowing a tail on $\widehat\phi(k)$. This is because
the unscattered tail of $\widehat\phi(k)$ could contribute as much to
scattering into $C_\Sigma$ as genuine scattering from near $k_0$. Such
pathological events correspond to situations in which the particle would
typically not be aimed at the target and in fact would not be detected at
all. The use of $T$ in (\ref{dT}), and $\Rightarrow$ in (\ref{dO}), has the
desirable effect of not counting such events.

It is shown by Amrein, Jauch, and Sinha \cite{AmreinJauch} that
\begin{equation}\label{avy}
\lim_{|\widehat\psi_{\rm in}(k)|^2  \to \delta(k-k_0)}\int_{C_{\Sigma}}\int_{y\in\Gamma_L}
\left|\widehat{T\psi_{{\rm in},y}}(k)\right|^2 \,{\rm d}^2y\, {\rm
d}^3k=16\pi^4\int_\Sigma |T(\omega|k_0|, k_0) |^2\,{\rm d}\Omega\,.
\end{equation} 
They compute
\begin{eqnarray}\label{avga}
\displaystyle
 \int_{a \perp k_0}
\left|\widehat{T\psi_{{\rm in},a}}(k)\right|^2 \,{\rm d}^2 a & = &
4\pi^2 \int_{a\perp k_0} \left|\int_{|k'|=|k|}
T(k,k') {\rm e}^{{\rm i}a\cdot k'}
\widehat \psi_{\rm in}(k') |k'|{\rm d}\Omega'\right|^2 \,{\rm d}^2a
\nonumber\\\nonumber\\ & & \hspace{-1cm} =\,\, 16 \pi^4\int_{|k'|=|k|}(\cos\theta')^{-1}
\left|T(k,k')\right|^2\left|\widehat\psi_{\rm in}(k') \right|^2\,{\rm d}\Omega'\,,
\end{eqnarray} 
where $\theta'$ is the angle between $k_0$ and $k'$. For the second
equality one uses that the $a$-integration over ${\rm e}^{{\rm
i}a\cdot(k'-k'')}$ produces $(2\pi)^2\delta(k'_\perp-k''_\perp)$, $k_\perp$
being the projection of $k$ on on the plane perpendicular to $k_0$. This in
turn yields effectively a $\delta(\omega'-\omega'')$ if one assumes that
$\widehat \psi_{\rm in}$ is supported in a neighborhood of $k_0$ that is
contained in the half space $P_{k_0} := \{ k\in\IR^3: k\cdot k_0 \geq
0\}$. Then in the limit $|\widehat\psi_{\rm in}(k)|^2 \to \delta(k-k_0)$
the r.h.s. of (\ref{avga}) becomes $16 \pi^4
\left|T(k,k_0)\right|^2\delta(|k|-|k_0|)$, and integrating this over
$C_{\Sigma}$ yields (\ref{avy}). (It is clear from the right hand side of
(\ref{avga}) that (\ref{avga}) is invariant under translations of
$\psi_{\rm in}$, so that (\ref{avy}) is independent of $L$.)

Writing for $\Omega_+^{-1}\phi_y$ in (\ref{dO}) 
\begin{equation}
\Omega_+^{-1}\phi_y=S\phi_y+\Omega_+^{-1}\phi_y-\Omega_+^{-1}\Omega_-\phi_y=T\phi_y+\phi_y+\Omega_+^{-1}(\phi_y-\Omega_-\phi_y)
\end{equation}
we see that  (\ref{diff}) 
then follows from the  condition
\begin{equation}\label{c}
\lim_{L\to\infty}\int_{y\in\Gamma_L}\|T(\phi_y-\Omega_-^{-1}\phi_y)\|^2\,{\rm
d}^2 y=0,
\end{equation}
which is presumably typically satisfied, although we are aware of no proof
of this. With (\ref{c}), we need only invoke (\ref{avy}) with $\psi_{\rm
in}=\phi$.

We remark that (\ref{c}) is considerably weaker than the simpler-looking
sufficient condition $\lim_{L\to\infty}\int_{y\in\Gamma_L}\|\Omega_- \phi_y
-\phi_y \|^2\,{\rm d}^2 y=0$: The application of $T$ may drastically
diminish $\phi_y-\Omega_-^{-1}\phi_y$. To appreciate this, note that as
$L\to\infty$, $T\psi_{\rm in}^y$ itself becomes very small.  As you
translate $\phi$ away from the scattering region, it has further to go
before it gets there. Thus, since wave functions spread under the (free)
time evolution, in all directions, when the wave function begins very far
away, it develops a large lateral spread by the time the scattering region
is approached and hence, since the scattering region is more or less
localized, most of the wave function does not scatter. We note also that it
is shown in \cite{thesis} that for a quite general class of short-range
potentials $\lim_{L\to\infty} \|\Omega_- \phi_y -\phi_y \| =0$ if
$|\widehat\phi(k)|^2 \approx \delta(k-k_0)$.\footnote{More generally, it is
shown \cite{thesis} that this result holds whenever $\phi$ is such that
$\widehat\phi$ is supported in the half space $P_{k_0}$. The proof of this
is very similar to the proof of the well known fact that the analogous
result holds for $\psi_L := {\rm e}^{{\rm i}LH_0}\psi$, i.e., when one
moves the state sufficiently far backwards in time according to the free
time evolution (see, e.g., \cite{Perry}).}

We wish to emphasize that the integration over the impact parameter, i.e.,
over $y$, is crucial not merely for the proof of (\ref{diff}) but for the
result itself. If all of the particles in the beam had the very same
initial wave function $\phi_L$, the total cross section---the integral of
the differential cross section over $S^2$---would then depend on detailed
geometrical characteristics of $\phi_L$ such as the impact parameter and
the distance $L$ to the target. Even if $\phi_L$ were an approximate plane
wave, with more or less constant modulus over most of its support, by the
time it had approached the target it would have developed a slowly varying
profile whose spread and whose position relative to the target would be
crucial for the total cross section.  Experimenters don't have to worry
much about such details because they work with homogeneous beams having a
random impact parameter.
  
\section{Naive scattering theory and the naive cross section}\label{nst}

The formula (\ref{T}) is not very concrete. How does one actually compute
$T$\,? Using heuristic stationary methods, this was first done by Max Born
\cite{Born} in the first paper on quantum mechanical scattering theory, in
which also the statistical law $\rho=|\psi|^2$ first appeared! We shall
review here how ``stationary scattering theory'' can be exploited to
rigorously obtain a formula for $T$ linking the stationary and the
time-dependent methods.

Consider solutions $\psi$ of the stationary Schr\"odinger 
equation with the asymptotics 
\begin{equation}\label{naive}
\psi(x) \approx {\rm e}^{{\rm i}k_0\cdot x} +
f^{k_0}(\omega)\frac{{\rm e}^{{\rm i}|k_0||x|}}{|x|}\,\quad {\rm for} 
\,|x|\,{\rm large}\,. 
\end{equation}
In naive scattering theory (cf., e.g., Notes to Chapter XI.6 in \cite{RS3})
the first term is regarded as representing an incoming 
plane wave and the second the outgoing scattered wave with angle-dependent
amplitude. 

Such wave functions  can be obtained as solutions of the Lippmann-Schwinger 
equation 
\begin{equation} \label{LS}
\psi(x,k) = {\rm e}^{{\rm i}k\cdot x} - \frac{1}{2\pi}\int \frac{ {\rm e}^{{\rm i}|k||x-y|}}{|x-y|}
V(y)\psi(y,k)\,{\rm d}^3y\,.
\end{equation}
These solutions form a complete set, in the sense that an expansion in
terms of these generalized eigenfunctions, a so-called generalized Fourier
transformation, diagonalizes the continuous spectral part of $H$. (In fact
from the intertwining relation (\ref{inter2}) one sees that $\psi(x,k) =
\langle x|\Omega_-|k\rangle$.) Hence the $T$-matrix can be expressed in
terms of generalized eigenfunctions and one finds (cf.\ \cite{RS3}) that
\begin{equation}\label{TV}
T(k,k') = (2\pi)^{-3} \int {\rm e}^{-{\rm i}k\cdot x} V(x) \psi(x,k')\,{\rm d}^3x\,.
\end{equation}
Thus the iterative solution of (\ref{LS}) yields  a perturbative expansion
for $T$, called the Born series.

Moreover, comparing (\ref{naive}) and (\ref{LS}),
expanding the right hand side of (\ref{LS}) in powers of $|x|^{-1}$, we see
from the leading term  that
\[
f^{k_0}(\omega) = -(2\pi)^{-1}\int {\rm e}^{-{\rm i}|k_0|\omega\cdot y} V(y) \psi(y,k_0)\, {\rm d}^3y\,.
\]
Thus $f^{k_0}(\omega) = -4\pi^2 T(\omega|k_0|, k_0)$.

In naive scattering theory, $f^{k_0}(\omega)$ is called the scattering
amplitude: One simply uses the stationary solutions of Schr\"odinger's
equation with the asymptotic behavior (\ref{naive}) to obtain the cross
section from the quantum probability flux through $\Sigma$ generated by the
scattered wave, suggesting the identification of the differential cross
section with
\begin{equation}\label{fcross} 
\sigma_{\rm naive}^{k_0}(\Sigma) : =
\int_\Sigma |f^{k_0}(\omega)|^2\,{\rm d}\Omega\,, 
\end{equation} 
in agreement with the result (\ref{diff}) sketched in the previous
section. However, such a heuristic derivation of the formula (\ref{fcross})
for the differential cross section, based solely on the stationary picture,
is unconvincing---even for physicists.

One can try to extract the time dependent picture  from the stationary one by
constructing wave packets from the generalized eigenfunctions $\psi(x,k)$;
see \cite{RS3}. Stationary phase ideas then suggest the development over
time of a transmitted and a scattered wave, corresponding to the two terms
in (\ref{naive}). However, unless the impact parameter is randomized, their
relative sizes---and hence the total cross section---will depend upon delicate cancellations contingent upon detailed geometrical considerations,
as indicated already at the end of Section~\ref{scs}.

\section{Scattering into cones: the cone cross section}\label{sic}

The analysis in Section~\ref{scs} is based on the formula (\ref{cross}) for
the scattering cross section, which is obtained by applying Born's
statistical law to momentum measurements in the distant future. But what
does the setup for scattering experiments, involving detectors covering
certain solid angles, have to do with the measurement of momentum? After
all, not every measurement is a momentum measurement. And in scattering
experiments each particle is ultimately detected at fairly definite
(though random) location---that of the detector that fires---after which
the state of the particle can hardly be regarded as a global plane wave,
which is what momentum measurements might reasonably be expected to
produce. If it is, in fact, appropriate to regard the final detection in a
scattering experiment as a measurement of momentum, it cannot be a priori
that this is so. Rather this must be justified by a quantum mechanical
analysis that takes the relevant experimental details into account.

These experimental details, involving detectors that locate particles at a
distant time in a given solid angle, suggest that the {\it cone cross
section\/}
\begin{equation}\label{conecross} 
\sigma^\psi_{\rm cone}(\Sigma)
:=\lim_{t\to\infty}\int_{C_\Sigma}|\psi_t(x)|^2\,{\rm d}^3x\, , 
\end{equation}
the asymptotic probability of finding the particle in the cone
$C_\Sigma$,\footnote{Note that $C_\Sigma$ in (\ref{conecross}) is the cone
in position space spanned by $\Sigma$.} is the more fundamental definition
of scattering cross section, more directly connected with what is measured
in a scattering experiment, and from which other formulas for the cross
section, such as (\ref{cross}), must be derived. This was accomplished by
Dollard \cite{Dollard} (see also \cite[p.\ 356]{RS3} and \cite{EnssSimon}),
whose scattering-into-cones theorem
\begin{equation}\label{SIC}
\lim_{t\to\infty}\int_{C_\Sigma}|\psi_t(x)|^2\,{\rm d}^3x =
\int_{C_\Sigma} |\widehat{\Omega_+^{-1}\psi}(k)|^2\,{\rm d}^3k
\end{equation}
says that $\sigma^\psi_{\rm cone}=\sigma^\psi$---that the cone cross
section is given by the simpler, more standard, though less fundamental
object (\ref{cross}).

\section{The flux cross section and the flux across surfaces theorem}\label{fas}
It is widely believed that the cone cross section (\ref{conecross}) more or
less directly conveys the statistics---the relative frequency of detector
firings---for the results of a scattering experiment.  But in a scattering
experiment does one actually determine whether the particle is in the cone
$C_\Sigma$ at some large  fixed time? Rather, is it not the case
that one of a collection of distant detectors, surrounding the scattering
center at a fairly definite distance, fires at some random time, a time
that is not chosen by the experimenter?  And isn't that random time simply
the time at which, roughly speaking, the particle crosses the surface of
the detector or detectors subtended by the cone? 

What a scattering experiment is fundamentally concerned with is not
scattering into cones but flux across surfaces.  Thus the quantum flux
$j^{\psi_t} = {\rm Im}\psi_t^*\nabla\psi_t$, the probability current for
the probability density $\rho_t(x)=|\psi_t(x)|^2$ in the quantum continuity
equation
\begin{equation}\label{continuity}
\frac{\partial \rho_t}{\partial t}+{\rm div}j^{\psi_t}=0\,,
\end{equation}
should play a fundamental role in scattering theory.  It is hard to resist
the suggestion that the quantum flux integrated over a surface gives the
probability that the particle crosses that surface, i.e., that
\begin{equation}\label{jdAdt}
j^{\psi_t}\cdot {\rm d}A{\rm d}t
\end{equation}
is the probability that a particle crosses the surface element ${\rm d}A$
in the time ${\rm d}t$. This suggestion must be taken ``cum grano salis''
since $j^{\psi_t}\cdot {\rm d}A{\rm d}t$ may somewhere be negative, in
which case it can't be a probability. However, in the scattering regime,
the regime we are interested in, this quantity is presumably positive far
away from the scattering center when ${\rm d}A$ is oriented outwards.

Hence, if the detectors are sufficiently distant from the scattering center
the flux will typically be outgoing and (\ref{jdAdt}) will be
positive,\footnote{In \cite{DDGZ2} the current positivity condition, which
states that the flux through a (given) surface is outgoing at all times,
was introduced.  In \cite{Leipzig} it is shown that this condition is
naturally associated with the dilation operator, whose spectral
decomposition is used in proving asymptotic completeness.}  so that it
appears natural to identify the probability that the particle crosses some
distant surface during some time interval, with the integral of
(\ref{jdAdt}) over that time interval and that surface. With this
identification, the integrated flux
provides us with a physically fundamental definition of the cross section:
\begin{equation}\label{fluxcross}
\sigma^\psi_{\rm flux}(\Sigma) := \lim_{R\to\infty}\int_0^\infty\, {\rm
d}t\,\int_{R\Sigma} j^{\psi_t}\cdot \,{\rm d}A\,, \end{equation} where
$R\Sigma$ is the intersection of the cone $C_\Sigma$ with the sphere of
radius $R$.  And a derivation of the formula (\ref{cross}) from microscopic
first principles then amounts to a proof of the {\it flux-across-surfaces
theorem}:
\begin{equation}\label{FAS} \lim_{R\to\infty}\int_0^\infty\, {\rm
d}t\,\int_{R\Sigma}j^{\psi_t}\cdot {\rm d}A =
\int_{C_\Sigma}|\widehat{\Omega_+^{-1}\psi}(k)|^2\,{\rm
d}^3k\,. \end{equation} The fundamental importance of the
flux-across-surfaces theorem was first recognized by Combes, Newton and
Shtokhamer \cite{CNS}. The first proof of the free flux-across-surfaces
theorem, i.e., for $V=0$, was given in \cite{FreeFlux}; a simplified
version of the proof can be found in \cite{Leipzig,thesis}. For proofs of
the flux-across-surfaces theorem for various classes of short and long
range potentials and under a variety of conditions on the wave function, see
\cite{AmreinZuleta,AmreinPearson,JMP}. (For more details on the proofs, we
refer the reader to the last section of this paper.)

Note that the flux-across-surfaces theorem (\ref{FAS}) also shows that the
scattering cross section (\ref{fluxcross}), defined via the quantum flux,
indeed yields a probability measure on the unit sphere. In fact, in the
course of establishing (\ref{FAS}) one also obtains that
\begin{equation}\label{FAS2}
\lim_{R\to\infty}\int_0^\infty\, {\rm d}t\,\int_{R\Sigma}j^{\psi_t}\cdot
{\rm d}A = \lim_{R\to\infty}\int_0^\infty\, {\rm
d}t\,\int_{R\Sigma}\left|j^{\psi_t}\cdot {\rm d}A\right|\,. \end{equation}
This shows that the flux is asymptotically outgoing and that the
identification of (\ref{jdAdt}) with the crossing probability is consistent
in the scattering regime.

\section{Random trajectories and the Bohmian cross section}\label{bcs}

There remains, however, a very serious difficulty with regarding the flux
cross section (\ref{fluxcross}) as the basic quantity for the derivation of
scattering theory from microscopic first principles, one that perhaps can
best be appreciated by asking: Precisely which microscopic principles have
been used for the derivation?  

Schr\"odinger's equation alone is certainly insufficient, since the
derivation involves quantum probability formulas and these transcend the
Schr\"odinger dynamics. A better answer would be standard textbook quantum
theory, involving, as well as Schr\"odinger's equation, the quantum
measurement postulates for the statistics of the results of
measurements of quantum observables. However, this theory, with the
macroscopic notion of measurement playing a {\it fundamental\/} role, is
not a fully microscopic theory and thus can't genuinely be regarded as
defining the microscopic first principles that we seek.

Moreover, even if we ignore this difficulty---as most physicists no doubt
would be inclined to do---there remains the severe difficulty that there is
no quantum observable, as understood in textbook quantum theory, to which
the quantum flux corresponds via the quantum measurement formalism. The
quantum flux is usually not regarded as having any operational
significance. It is not related to any standard quantum mechanical
measurement in the way, for example, that the density $\rho$, as the
spectral measure of the position operator, gives the statistics for a
position measurement.

We have proposed that the (time-integrated) flux be identified with a
crossing probability, the probability that the particle crosses a given
piece of surface---which, as we have emphasized, to the extent that we are
allowed to use such concepts at all in orthodox quantum theory, it does at
a random time. Thus the relevant observable should be the position of the
particle at a random time, the time at which it crosses the surface. This
time should, in orthodox quantum theory, be associated with a
time-operator. But the notion of time-operator is exceedingly
problematical, and the notion of the position at this random time is
utterly hopeless from an orthodox perspective.

There is, however, a suitable candidate for a theory embodying the
appropriate first principles, namely, Bohmian mechanics \cite{Bohm52,qe,e},
which provides a rigorous foundation for the ``suggestions'' and ``natural
identifications'' of Section~\ref{fas}.  In Bohmian mechanics a particle
moves along a trajectory $X(t)$ determined by (using now general units)
\begin{equation}\label{Bohm}
\frac{\rm d}{{\rm d}t}X(t) = v^{\psi_t}(X(t)) = \frac{\hbar}{m} {\rm Im}
\frac{\nabla\psi_t}{\psi_t}(X(t))\,,
\end{equation}
where $\psi_t$ is the particle's wave function, evolving 
according to Schr\"odinger's equation. Moreover, if an 
ensemble of particles with wave function $\psi$ is prepared, 
the positions $X$ of the particles are distributed according 
to the quantum equilibrium distribution $\IP^\psi$ with density 
$\rho=|\psi|^2$. 

In particular, since $|\psi_t|^2v^{\psi_t} = j^{\psi_t}$, the continuity
equation for the probability shows that the probability flux
$(|\psi_t|^2,|\psi_t|^2v^{\psi_t})$ is conserved, i.e., the flow
(\ref{Bohm}) carries an initial $|\psi|^2$ probability density for the
particle to the density $|\psi_t|^2$ at time $t$.  Thus, given an initial
wave function $\psi$, the solutions $X^\psi(t)\equiv X^\psi(t,X_0)$ of equation
(\ref{Bohm}) are random trajectories, with $X^\psi(t)$ having distribution
$|\psi_t(x)|^2$, and where the randomness comes from that of the
$\IP^\psi$-distributed initial position $X_0$.

Let now $\Sigma$ be any smooth piece of oriented surface in $\IR^3$ and
consider the number $N^\psi(\Sigma,I)$ of crossings by the trajectory $X^\psi(t)$ of
$\Sigma$ in the time interval $I$. Consider also $N^\psi_+(\Sigma,I)$, the
number of crossings in the direction of the orientation, and
$N^\psi_-(\Sigma,I)$, the number of crossings in the opposite direction, of
$\Sigma$ in the time interval $I$. Then
$N^\psi(\Sigma,I)=N^\psi_+(\Sigma,I)+N^\psi_-(\Sigma,I)$ and we define the number of
signed crossings by $N^\psi_{\rm s}(\Sigma,I) := N^\psi_+(\Sigma,I)-N^\psi_-(\Sigma,I)$.

We now compute the expectation values with respect to the probability
$\IP^\psi$ of these random variables in the usual manner. For a
crossing of an infinitesimal surface element of (vector) size ${\rm d}A$ to
occur in the time interval $(t,t+{\rm d}t)$, the particle must be in a
cylinder of size $|v^{\psi_t} {\rm d}t\cdot {\rm d}A|$ at time $t$. Thus
$\IE^\psi(N^\psi({\rm d}A,{\rm d}t)) = |\psi_t|^2|v^{\psi_t}{\rm
d}t\cdot {\rm d}A| = |j^{\psi_t}\cdot {\rm d}A|\,{\rm d}t$, and similarly
$\IE^\psi(N^\psi_{\rm s}({\rm d}A,{\rm d}t)) = j^{\psi_t}\cdot {\rm
d}A\,{\rm d}t$. Hence
\begin{equation}\label{n}
\IE^\psi(N^\psi(\Sigma,I))=\int_I \, \int_\Sigma |j^{\psi_t}\cdot {\rm d}A|\,{\rm d}t
\end{equation} 
and
\begin{equation}\label{ns}
\IE^\psi(N^\psi_{\rm s}(\Sigma,I))=\int_I \, \int_\Sigma j^{\psi_t}\cdot {\rm
d}A\,{\rm d}t\,.
\end{equation}

Consider now a particle with wave function $\psi$ localized, say, at time
$t=0$ in some region $B\subset\IR^3$ with smooth boundary $\partial B$.
The random variables $t^\psi_B$, the first exit time from $B$, $t^\psi_B:=
\inf\{t\geq 0\,|\,X^\psi(t)\notin B\}$, and $X^\psi_B$, the position of
first exit, $X^\psi_B := X^\psi(t_B)$, are the basic quantities describing
the exit of the particle from $B$. If $j^{\psi_t}\cdot {\rm d}A$
is, for all $t>0$, positive everywhere on $\partial B$, the particle can
cross $\partial B$ at most once and only outwards.   We then
have that for $\Sigma\subset\partial B$
\begin{equation}
\IP^\psi(X^\psi_B \in \Sigma) =
\IE^\psi(N^\psi_{\rm s}(\Sigma))\,, 
\end{equation}
where we have written $N^\psi_{\rm s}(\Sigma)$ for $N^\psi_{\rm
s}(\Sigma,(0,\infty))$, with a similar notation for $N^\psi$ and
$N^\psi_\pm$. More generally, since $|I_{\{X^\psi_B \in
\Sigma\}}-N^\psi_{\rm s}(\Sigma)|\leq N^\psi_-(\partial
B)=\frac12(N^\psi(\partial B)-N^\psi_{\rm s}(\partial B))$, where
$I_{\{\cdot\}}$ is the indicator function of $\{\cdot\}$, we have that
\begin{equation}\label{xn}
|\IP^\psi(X^\psi_B \in \Sigma) - \IE^\psi(N^\psi_{\rm
s}(\Sigma))|\leq\frac12\left(\IE^\psi(N^\psi(\partial B)) -\IE^\psi(N^\psi_{\rm s}(\partial B))\right)\,.
\end{equation}
We now define the Bohmian cross section as the probability that the
particle crosses the surface covered by the relevant detector or detectors
at some future time. More precisely, we define the Bohmian cross section
as the $R\to\infty$ limit of the probability that the particle will leave
the ball $B=B_R$, of radius $R$ centered at the origin, through $R\Sigma$,
$\Sigma\subset S^2$,
\begin{equation} \label{Bohmcross}
\sigma^\psi_{\rm Bohm} (\Sigma) := \lim_{R\to\infty} \IP^\psi(X^\psi_{B_R} \in R\Sigma)\,.
\end{equation} 
This is physically the most fundamental definition of the cross section,
corresponding more or less directly to what is measured in a scattering
experiment. This definition involves a quantity, the first exit position
$X^\psi_{B_R}$, which, while perfectly straightforward for Bohmian
mechanics, cannot  be expressed in orthodox quantum theory. 

It follows from (\ref{FAS2}) and (\ref{n}--\ref{xn}) that
$\sigma^\psi_{\rm Bohm}=\sigma^\psi_{\rm flux}$.

\section{Overview}\label{p}
Using (\ref{Bohmcross}) instead of (\ref{cross}) in the analysis leading to (\ref{dO}), we arrive at  
\begin{equation}\label{dc}
\sigma^{k_0}_{\rm diff}(\Sigma):=\lim_{|\hat\phi(k)|^2\Rightarrow\delta(k-k_0)}\lim_{L\to\infty}\int_{y\in\Gamma_L} \lim_{R\to\infty}\IP^{\phi_y}(X^{\phi_y}_{B_R} \in R\Sigma)\,{\rm d}^2y\,,
\end{equation} 
for $k_0\notin C_{\Sigma}$, as the fundamental definition of the
differential scattering cross, describing the scattering rate for a beam of
particles of momentum $k_0$.  Our derivation of scattering theory from
microscopic first principles thus becomes the demonstration from Bohmian
mechanics of the emergence of (\ref{diff}) from (\ref{dc}). It is worth
noting that (\ref{dc}) is somewhat complicated, involving three explicit
limits, each crucial and with the order of the limits important. For
example, because the limit $R\to\infty$ is taken first, the wave functions
$\phi_y$ are asymptotically in the support of $B_R$. 

The derivation begins with the analysis of Section~\ref{bcs} and proceeds
via the flux-across-surfaces theorem, (\ref{FAS}) and (\ref{FAS2}), to
(\ref{dO}). Then, using the computation of Amrein, Jauch, and Sinha
described in Section~\ref{scs}, we arrive at (\ref{diff}), which in turn
can be computed using the stationary methods described in
Section~\ref{nst}. One of the frequent objections against Bohmian mechanics
is that it lacks the resources to cope, e.g., with momentum, based as it is
solely upon position. It is thus worth emphasizing that our analysis shows
how the usual textbook scattering formulas involving momentum matrix
elements naturally emerge from Bohmian mechanics.

We wish to comment now on a crucial step in the derivation: the
flux-across-surfaces theorem. Note that there is a peculiarity in the
statement of that theorem: The right hand side of (\ref{FAS}) is well
defined for all wave functions in the range of $\Omega_+$, but one cannot
expect the theorem to hold for all such wave functions because the left
hand side, involving the flux, is defined only if the wave function obeys
certain smoothness conditions. 

The usual mathematical physics of scattering theory, with its focus on
asymptotic completeness, neither relies upon nor needs such smoothness
properties, nor does Dollard's theorem (\ref{SIC}), but to treat the flux,
extra conditions and new techniques are required. One might expect that
(\ref{FAS}) holds whenever the wave functions are sufficiently smooth and
are moving freely asymptotically in time, i.e., are in the range of
$\Omega_+$.  But this has not yet been shown! One typical problem, for
example, is that the standard techniques in time-dependent scattering
theory yield the required ``propagation estimates'' only for wave functions
with energy cutoffs for small and large energies (cf.\
\cite{AmreinZuleta,AmreinPearson}).  When proving asymptotic completeness,
these are harmless  because they can be easily removed at the
appropriate time by simple density-in-$L^2$ arguments. However, this does not
work in (\ref{FAS}) because of the unboundedness of the form $\int_0^\infty
\,{\rm d}t \, \int_{R\Sigma} j^{\psi_t}\cdot {\rm d}A$. On the other hand,
the few known propagation estimates for wave functions without energy
cutoffs (cf.\ \cite{Soffer,Yajima}) are not strong enough for proving the
flux-across-surfaces theorem.

One way to come to grips with this is to turn to generalized eigenfunction
expansions (see \cite{Ikebe,JMP,Panati}). However, while no energy cutoffs
are then needed, the class of allowed potentials in \cite{JMP} is less general
than in the standard approaches \cite{AmreinZuleta,AmreinPearson}.
Nevertheless, the eigenfunction expansions have proven to be a general and
rather promising tool. Further mathematical work on generalized
eigenfunctions would surely be of interest for the foundations of
scattering theory. We recall in this respect also the use of (\ref{LS}) for
actual computation.

It would be very interesting to know whether the 
energy cutoffs on the wave functions can be circumvented 
without sacrificing the less restrictive conditions on the potential
appearing in the standard approaches to the proof of (\ref{FAS}).
As mentioned before, the most general and most satisfying result would be
that any sufficiently smooth wave function whose motion is
asymptotically free, i.e., that is in the range of $\Omega_+$, 
satisfies (\ref{FAS}). This would justify the name scattering
states for the set ${\rm Ran}(\Omega_+)$. On the other hand it would be 
interesting to understand whether (\ref{Bohmcross}) is a well defined
probability measure also for states in the singular continuous 
spectral subspace, even though the formula (\ref{FAS}) could then no longer
hold.  

For the case of many-particle scattering, asymptotic completeness has been
established by Soffer and Sigal (see \cite{Dere,SigalSoffer} and the
references therein). Moreover, Bohmian mechanics for many-particle systems
is perfectly well defined \cite{e}. However, we are not aware of any work
on a many-particle analogue of the flux-across-surfaces theorem, which
would be necessary for a more complete understanding of many-particle
scattering phenomena in terms of microscopic first principles.

\end{document}